\documentclass{Interspeech2024}
\usepackage{color,amsmath,url,times,tabularx,bbm,amssymb,multirow}
\usepackage{footnote}
\makesavenoteenv{tabular}
\makesavenoteenv{table}
\usepackage{makecell} 
\usepackage{cite}

\usepackage{graphicx}
\usepackage{bm}
\usepackage{comment}

\interspeechcameraready

\title{DiffATR: Diffusion-based Generative Modeling for Audio-Text Retrieval}

\name[affiliation={}]{Yifei}{Xin}
\name[affiliation={}]{Xuxin}{Cheng}
\name[affiliation={}]{Zhihong}{Zhu}
\name[affiliation={}]{Xusheng}{Yang}
\name[affiliation={*}]{Yuexian}{Zou}

\address{School of ECE, Peking University, China \thanks{This paper was partially supported by NSFC (No: 62176008) and Shenzhen Science \& Technology Research Program (No:GXWD20201231165807007-20200814115301001).}\thanks{$^{*}$ Yuexian Zou is the corresponding author.}}
\email{xinyifei@stu.pku.edu.cn}


\begin{document}

\maketitle
\begin{abstract}
Existing audio-text retrieval (ATR) methods are essentially discriminative models that aim to maximize the conditional likelihood, represented as $p(candidates|query)$. Nevertheless, this methodology fails to consider the intrinsic data distribution $p(query)$, leading to difficulties in discerning out-of-distribution data. In this work, we attempt to tackle this constraint through a generative perspective and model the relationship between audio and text as their joint probability $p(candidates,query)$. To this end, we present a diffusion-based ATR framework (DiffATR), which models ATR as an iterative procedure that progressively generates joint distribution from noise. Throughout its training phase, DiffATR is optimized from both generative and discriminative viewpoints: the generator is refined through a generation loss, while the feature extractor benefits from a contrastive loss, thus combining the merits of both methodologies. Experiments on the AudioCaps and Clotho datasets with superior performances, verify the effectiveness of our approach. Notably, without any alterations, our DiffATR consistently exhibits strong performance in out-of-domain retrieval settings. 
\end{abstract}

\section{Introduction}
Given a query in the form of an audio clip or a caption, audio-text retrieval (ATR) seeks to identify a matching item from a collection of candidates in another modality. Existing ATR techniques \cite{koepke2022audio,mei2022metric,xin23c_interspeech} predominantly interpret cross-modal interactions via discriminative modeling based on contrastive learning. However, from a probabilistic viewpoint, discriminative models \cite{nallapati2004discriminative} only capture the conditional probability distribution, denoted as $p(candidates|query)$. This leads to a limitation that discriminative models fail to model the underlying data distribution \cite{bernardo2007generative}, and their latent space includes fewer intrinsic data characteristics $p(query)$, which challenges their ability to generalize on unseen data \cite{liang2022gmmseg}. In contrast, generative models capture the joint probability distribution between a query and candidates as $p(candidates,query)$, ensuring a more semantic-based latent mapping based on the data's inherent characteristics. As a result, generative models show superior generalizability and transferability compared to the discriminative counterparts. 

Recent advancements in generative modeling have been notable, with breakthroughs in image synthesis \cite{dhariwal2021diffusion}, natural language processing \cite{brown2020language}, and audio generation \cite{liu2023audioldm}. Among the leading generative techniques is the diffusion model \cite{yang2022diffusion,ho2020denoising}, which is a type of likelihood-based approach that refines signals by iteratively eliminating noise through a learned denoising model. The diffusion model's unique ability to refine from coarse to fine-grained stages makes it adept at capturing the relationship between audio and text. With this in mind, we employ the diffusion model to represent the audio-text correlation via their joint probability.

To this end, we introduce DiffATR, a diffusion-based ATR framework, to tackle the shortcomings of discriminative approaches from a generative perspective. We attempt to model the ATR task as an iterative procedure that progressively crafts a joint distribution from noise. Given a query and a collection of candidates, we adopt the diffusion model to generate the joint probability distribution $p(candidates,query)$. To enhance the ATR performance, we optimize our method from both generative and discriminative angles. During the training stage, the generator is refined using a common generation loss, i.e., Kullback-Leibler (KL) divergence \cite{kullback1997information}, while the feature encoder is optimized by the contrastive loss, i.e., NT-Xent loss \cite{chen2020simple}, ensuring a blend of discriminative accuracy and generative flexibility. 

DiffATR presents two key strengths. Firstly, its generative nature makes it inherently generalizable and adaptable to out-of-domain samples without additional adjustments. Secondly, the iterative refinement property of the diffusion model allows DiffATR to progressively enhance the retrieval results from coarse to fine. Experimental results on two benchmark ATR datasets, including AudioCaps and Clotho, demonstrate the advantages of DiffATR. To further evaluate the generalization of our method to unseen data, we introduce an out-of-domain ATR task, where the goal is to retrieve relevant audio/text data from a dataset that belongs to a domain different from the one the model is originally trained on. Our method not only represents a novel effort to promote generative approaches for in-domain ATR, but also shows the potential of generative solutions in out-of-domain ATR settings. 

The primary contributions of this paper include:
\begin{itemize} 
\item To the best of our knowledge, we are the first to boost ATR task from a generative viewpoint. We introduce a novel diffusion-based ATR framework (DiffATR) that models ATR as an iterative procedure to progressively generate joint distribution from noise.
\item DiffATR effectively improves the retrieval performance on multiple ATR benchmarks and can generalize to different baseline models.
\item More encouragingly, DiffATR consistently excels in out-of-domain ATR without extra adjustments.
\end{itemize}

\begin{figure}[t]
  \centering
  \includegraphics[width=1.0\linewidth]{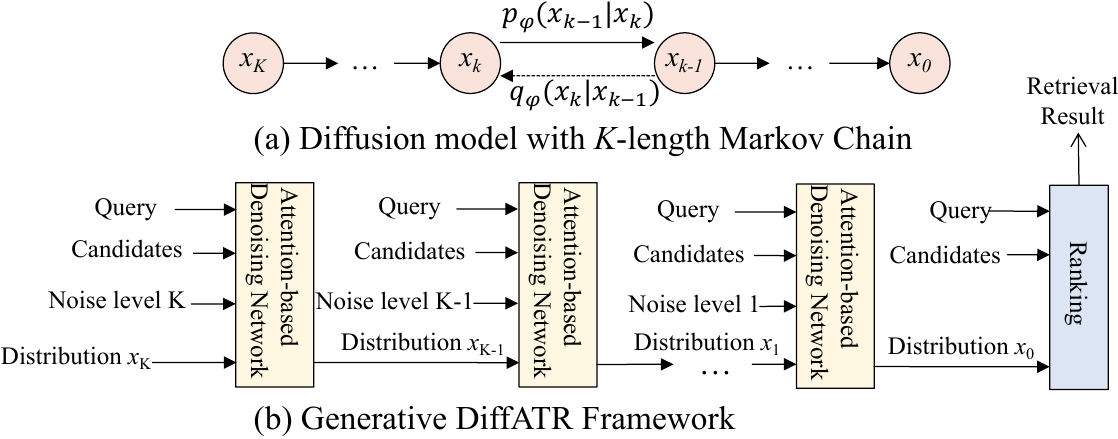}
  \caption{(a) The diffusion model for joint probability generation. (b)  Overview of our DiffATR framework.}
  \label{fig:adapt-pic}
  \vspace*{-\baselineskip}
\end{figure}

\section{Proposed Method}	
\subsection{Existing Discriminative Modeling}	
For ATR methods, similarity learning \cite{li2021multi} is often employed. Typically, both text and audio embeddings are mapped to a shared latent space, where their similarity is calculated as the dot product of their corresponding representations. In text-to-audio retrieval, such approaches \cite{mei2022metric} utilize logits to calculate the posterior probability:	
\begin{equation}
p\left(a|t ; \boldsymbol{\theta}_{\boldsymbol{t}}, \boldsymbol{\theta}_{\boldsymbol{a}}\right)=\frac{\exp \left(f_{\boldsymbol{\theta}_{\boldsymbol{t}}}(t)^{\top} f_{\boldsymbol{\theta}_{\boldsymbol{a}}}(a) / \tau\right)}{\sum_{a^{\prime} \in \boldsymbol{A}} \exp \left(f_{\boldsymbol{\theta}_t}(t)^{\top} f_{\boldsymbol{\theta}_{\boldsymbol{a}}}\left(a^{\prime}\right) / \tau\right)},
\end{equation}	
where $\tau$ is the temperature hyper-parameter, $\boldsymbol{A}$ represents the corpus of audio clips. $\boldsymbol{\theta}_{\boldsymbol{t}}$ and $\boldsymbol{\theta}_{\boldsymbol{a}}$ denote the text and audio encoder parameters, respectively. Then, existing approaches rank all audio candidates based on the posterior probability $p\left(a|t ; \boldsymbol{\theta}_{\boldsymbol{t}}, \boldsymbol{\theta}_{\boldsymbol{a}}\right)$. Similarly, in audio-to-text retrieval, they rank all text candidates based on $p\left(t|a ; \boldsymbol{\theta}_{\boldsymbol{t}}, \boldsymbol{\theta}_{\boldsymbol{a}}\right)$. The parameters $\left\{\boldsymbol{\theta}_{\boldsymbol{t}}, \boldsymbol{\theta}_{\boldsymbol{a}}\right\}$ of text and audio encoders are optimized by minimizing the contrastive loss:
\begin{equation}
\begin{array}{r}
\boldsymbol{\theta}_{\boldsymbol{t}}^*, \boldsymbol{\theta}_{\boldsymbol{a}}^*=\underset{\boldsymbol{\theta}_{\boldsymbol{t}}, \boldsymbol{\theta}_{\boldsymbol{a}}}{\arg \min }-\frac{1}{2} \mathbb{E}_{(t, a) \in \mathcal{D}}\left[\log p\left(a|t ; \boldsymbol{\theta}_{\boldsymbol{t}}, \boldsymbol{\theta}_{\boldsymbol{a}}\right)\right. \\
\left.+\log p\left(t|a ; \boldsymbol{\theta}_{\boldsymbol{t}}, \boldsymbol{\theta}_{\boldsymbol{a}}\right)\right]
\end{array}
\end{equation}
where $\mathcal{D}$ is a collection of text-audio pairs $(t, a)$. Such learning method is equivalent to maximizing conditional likelihood, i.e., $\prod_{(t, a) \in \mathcal{D}} p(a|t)+\prod_{(t, a) \in \mathcal{D}} p(t|a)$, which is called discriminative training \cite{kaur2021comparative}.

Since current approaches focus solely on the conditional probability distribution $p(a|t)+p(t|a)$, without considering the input distribution $p(t)$ and $p(a)$, they fail to generalize well to unseen and out-of-domain data.

\subsection{DiffATR: Generation Modeling}
As shown in Figure 1, we approach the ATR task from a generative modeling perspective. Drawing inspiration from the application of diffusion models \cite{zhang2023irgen,jin2023diffusionret}, we employ the diffusion model as the generator. Specifically, given a query and $N$ candidates, we aim to generate the distribution $x^{1: N}=\left\{x^i\right\}_{i=1}^N$ from Gaussian noise $\mathcal{N}(0, \mathrm{I})$. In contrast to the prior works that primarily focus on optimizing the posterior probabilities $p\left(a|t ; \boldsymbol{\theta}_{\boldsymbol{t}}, \boldsymbol{\theta}_{\boldsymbol{a}}\right)+$ $p\left(t|a ; \boldsymbol{\theta}_{\boldsymbol{t}}, \boldsymbol{\theta}_{\boldsymbol{a}}\right)$, our approach constructs the joint probabilities:
\begin{equation}
x^{1: N}=p(a, t|\phi)=f_\phi(a, t, \mathcal{N}(0, \mathrm{I}))
\end{equation}
where $\phi$ represents the generator's parameter. Notably, the generator's learning objective is equivalent to approximating the data distribution, i.e., $\prod_{(t, a) \in \mathcal{D}} p(a, t)$, which is termed generative training \cite{bernardo2007generative}. 

\begin{figure}[t]
  \centering
  \includegraphics[width=1.0\linewidth]{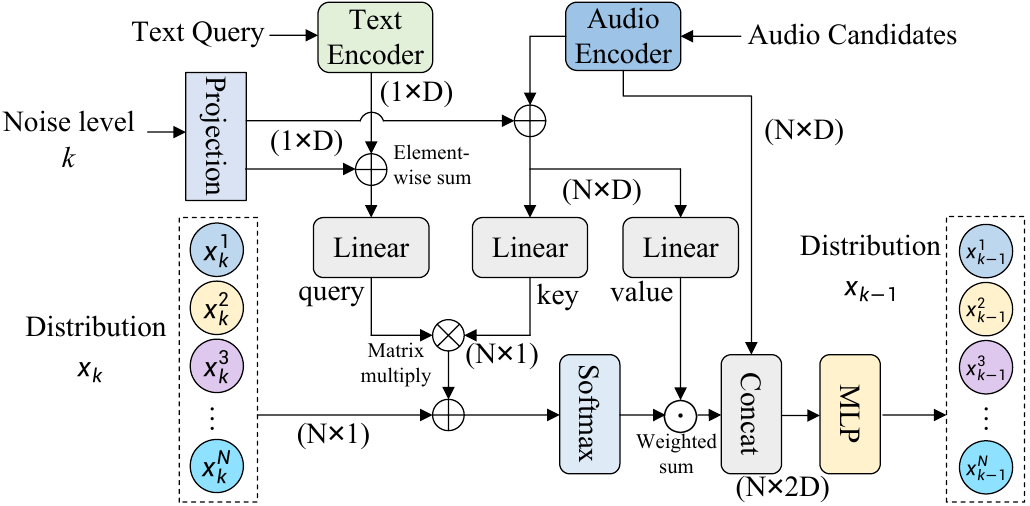}
  \caption{Model Architecture of the Attention-based Denosing Module.}
  \vspace*{-\baselineskip}
\end{figure}

\begin{table*}
  \caption{Performance comparison of our proposed method with baseline methods on the AudioCaps dataset.}
  \vspace*{-\baselineskip}
  \centering
  \begin{tabular}{c|cccccc}
    \toprule
    \multirow{3}{*}{Methods}
    & \multicolumn{6}{c}{AudioCaps}\\
    \cline{2-7}
    & \multicolumn{3}{c}{Text-to-Audio} & \multicolumn{3}{c}{Audio-to-Text}\\
    \cline{2-7}
    & \textbf{R@1} & \textbf{R@5} & \textbf{R@10} & \textbf{R@1} & \textbf{R@5} & \textbf{R@10}\\
    \midrule
    ResNet38 (baseline) \cite{mei2022metric} & 33.9$\pm$0.4 & 69.7$\pm$0.2 & 82.6$\pm$0.3 & 39.4$\pm$1.0 & 72.0$\pm$1.0 & 83.9$\pm$0.6\\
    CNN14 (baseline) \cite{mei2022metric} & 31.4$\pm$0.4 & 66.2$\pm$0.3 & 78.8$\pm$0.2 & 38.2$\pm$0.5 & 68.7$\pm$0.5 & 81.9$\pm$0.3\\
    ResNet38 (Diff) & \textbf{36.1$\pm$0.2} & \textbf{71.9$\pm$0.5} & \textbf{84.9$\pm$0.1} & \textbf{42.6$\pm$0.3} & \textbf{74.4$\pm$0.4} & \textbf{86.6$\pm$0.6}\\
    CNN14 (Diff) & \textbf{34.3$\pm$0.3} & \textbf{68.8$\pm$0.3} & \textbf{81.4$\pm$0.5} & \textbf{41.1$\pm$0.2} & \textbf{71.4$\pm$0.6} & \textbf{84.9$\pm$0.7}\\   
  \bottomrule
\end{tabular}
\vspace*{-\baselineskip}
\end{table*}

\subsubsection{Attention-based Denoising Module}
Different from general audio generation tasks which mainly prioritize the authenticity and diversity of generated samples, the key to ATR task lies in mining the correspondence between the query and candidates. To this end, we introduce an attention-based denoising module to capture the relationship between query and candidates during the generation process. The overview of DiffATR's denoising module is shown in Figure 2.

For the text-to-audio retrieval task, we first project the text representation $C_t \in \mathbb{R}^D$ extracted from the text encoder into the query and the audio representation $C_a \in \mathbb{R}^{N \times D}$ from the audio encoder into key and value, where $N$ denotes the number of audio candidates and $D$ is the number of feature channels. The projections are represented as: $Q_t=W_{Q_t}\left(C_t+pj(k)\right)$, $K_a=W_{K_a}\left(C_a+pj(k)\right)$, $V_a=W_{V_a}\left(C_a+pj(k)\right)$, where $W_{Q_t}, W_{K_a}$ and $W_{V_a}$ are projection matrices. The projection function $pj(\cdot)$ transforms the noise level $k$ into $D$ dimensional embedding. To give more weights to the audio candidates with higher joint probabilities of the previous noise level, we add the distribution $x_k$ to the attention weight. The attention mechanism is formulated as: $E_t=\operatorname{Softmax}\left(Q_t K_a^{\top}+x_k\right) V_a$.

The attention module's output $E_t$ serves as a high-level semantic embedding infused with textual query information. Then, we concatenate the audio representation $C_a$ and the embedding $E_t$ as the input of the denoising decoder $\left[C_a, E_t\right] \in \mathbb{R}^{N \times 2 D}$. The denoising decoder is a multi-layer perceptron (MLP) including two linear layers with a Relu activation function to compute the output distribution.

Similarly, for audio-to-text retrieval, we use the projected audio representation as query and the projected text representation as key and value into our attention-based denoising module. The output distribution is computed in the same way.

\begin{table*}
  \caption{Performance comparison of our proposed method with baseline methods on the Clotho dataset.}
  \vspace*{-\baselineskip}
  \centering
  \label{tab:freq}
  \begin{tabular}{c|cccccc}
    \toprule
    \multirow{3}{*}{Methods}
    & \multicolumn{6}{c}{Clotho} \\
    \cline{2-7}
    & \multicolumn{3}{c}{Text-to-Audio} & \multicolumn{3}{c}{Audio-to-Text} \\
    \cline{2-7}
    & \textbf{R@1} & \textbf{R@5} & \textbf{R@10} & \textbf{R@1} & \textbf{R@5} & \textbf{R@10}\\
    \midrule
    ResNet38 (baseline) \cite{mei2022metric} & 14.4$\pm$0.4 & 36.6$\pm$0.2 & 49.9$\pm$0.2 & 16.2$\pm$0.7 & 37.5$\pm$0.9 & 50.2$\pm$0.7\\
    CNN14 (baseline) \cite{mei2022metric} & 13.9$\pm$0.3 & 34.3$\pm$0.5 & 48.2$\pm$0.8 & 14.3$\pm$0.8 & 35.9$\pm$0.6 & 49.9$\pm$0.5\\
    ResNet38 (Diff) & \textbf{16.7$\pm$0.1} & \textbf{38.2$\pm$0.3} & \textbf{51.9$\pm$0.1} & \textbf{18.8$\pm$0.2} & \textbf{40.4$\pm$0.5} & \textbf{52.7$\pm$0.3}\\
    CNN14 (Diff) & \textbf{16.1$\pm$0.6} & \textbf{36.6$\pm$0.1} & \textbf{50.3$\pm$0.2} & \textbf{17.2$\pm$0.3} & \textbf{39.4$\pm$0.4} & \textbf{51.6$\pm$0.5}\\   
  \bottomrule
\end{tabular}
\vspace*{-\baselineskip}
\end{table*}

\subsubsection{Generation-based Optimization}
From the generation viewpoint, we model the distribution $x=p(a, t|\phi)$ as the reversed diffusion process of a Markov Chain with length $K$. We learn the joint distribution of text and audio by iteratively denoising a variable sampled from Gaussian distribution. In a forward diffusion process $q\left(x_k|x_{k-1}\right)$, the noise sampled from Gaussian distribution is added to a ground truth data distribution $x_0$ at every noise level $k$:
\begin{equation}
\begin{aligned}
& q\left(x_k|x_{k-1}\right)=\mathcal{N}\left(x_k ; \sqrt{1-\beta_k} x_{k-1}, \beta_k \mathrm{I}\right), \\
& q\left(x_{1: K}|x_0\right)=\prod_{k=1}^K q\left(x_k|x_{k-1}\right),
\end{aligned}
\end{equation}
where $\beta_k$ decides the step size which gradually increases, the ground truth vector $x_0$ is defined such that only the candidate which matters the query is one, others are all zeros (e.g., $\left[1,0,0,...,0\right]$). We can sample $x_k$ by the following formula:
\begin{equation}
x_k=\sqrt{{\alpha}^{\ast}_k} x_0+\sqrt{1-{\alpha}^{\ast}_k} \epsilon_k,
\end{equation}
where ${\alpha}^{\ast}_k=\prod_{i=1}^k \alpha_i$ and $\alpha_k=1-\beta_k . \epsilon_k$ is a noise sampled from $\mathcal{N}(0, \mathrm{I})$. We follow \cite{ramesh2022hierarchical} to predict the data distribution itself, i.e., $\hat{x}_0=f_\phi\left(a, t, x_k\right)$. The training objective of the diffusion model is formulated as:
\begin{equation}
\mathcal{L}_G=\mathbb{E}_{(t, a) \in \mathcal{D}}\left[\operatorname{KL}\left(\hat{x}_0 \vert x_0\right)\right]
\end{equation}
This loss maximizes the likelihood of $p(a, t)$ by bringing $f_\phi\left(a, t, x_k\right)$ and $x_0$ closer together.

\begin{table*}
  \caption{Text-to-audio and audio-to-text retrieval performances in out-of-domain retrieval settings. “Clotho $\rightarrow$ AudioCaps” indicates that “Clotho” is the source domain (the training dataset) and “AudioCaps” is the target domain (the evaluation dataset).}
  \vspace*{-\baselineskip}
  \centering
  \begin{tabular}{c|cccccc}
    \toprule
    \multirow{3}{*}{Methods}
    & \multicolumn{6}{c}{Clotho $\rightarrow$ AudioCaps}\\
    \cline{2-7}
    & \multicolumn{3}{c}{Text-to-Audio} & \multicolumn{3}{c}{Audio-to-Text} \\
    \cline{2-7}
    & \textbf{R@1} & \textbf{R@5} & \textbf{R@10} & \textbf{R@1} & \textbf{R@5} & \textbf{R@10}\\
    \midrule
    ResNet38 (baseline) & 12.4$\pm$0.3 & 35.8$\pm$0.5 & 50.6$\pm$0.4 & 15.0$\pm$0.2 & 39.9$\pm$0.6 & 52.1$\pm$0.8\\
    CNN14 (baseline) & 11.5$\pm$0.6 & 34.9$\pm$0.5 & 48.8$\pm$0.3 & 12.2$\pm$0.9 & 36.4$\pm$0.8 & 49.9$\pm$0.4\\
    ResNet38 (Diff) & \textbf{17.2$\pm$0.2} & \textbf{40.2$\pm$0.3} & \textbf{54.3$\pm$0.2} & \textbf{19.4$\pm$0.3} & \textbf{44.3$\pm$0.6} & \textbf{56.6$\pm$0.2}\\
    CNN14 (Diff) & \textbf{15.4$\pm$0.3} & \textbf{39.4$\pm$0.7} & \textbf{53.3$\pm$0.8} & \textbf{16.8$\pm$0.1} & \textbf{40.4$\pm$0.6} & \textbf{54.1$\pm$0.4}\\   
  \bottomrule
\end{tabular}
\vspace*{-\baselineskip}
\end{table*}

\subsubsection{Discrimination-based Optimization} 
From the discrimination perspective, we leverage the contrastive learning method to optimize the audio/text features so that these features hold discriminative semantic information. For a text-audio pair $(t,a)$, the similarity of the text and audio is measured by the cosine similarity of their embeddings:
\begin{equation}
             s(t, a) = \frac{f_{\boldsymbol{\theta}^{\boldsymbol{t}}}(t)^{\top} \cdot f_{\boldsymbol{\theta}^{\boldsymbol{a}}}(a)}{{\Vert f_{\boldsymbol{\theta}^{\boldsymbol{t}}}(t) \Vert}_2 {\Vert f_{\boldsymbol{\theta}^{\boldsymbol{a}}}(a) \Vert}_2}.
\end{equation}
Then, the contrastive loss can be formulated as:
\begin{equation}
\begin{aligned}
\mathcal{L}_D=-\frac{1}{2} \mathbb{E}_{(t, a) \in \mathcal{D}}&\left[\log \frac{\exp \left(s(t, a) / \tau\right)}{\sum_{a^{\prime} \in \boldsymbol{A}} \exp \left(s(t, a^{\prime}) / \tau\right)}\right. \\
&\left.+\log \frac{\exp \left(s(t, a) / \tau\right)}{\sum_{t^{\prime} \in \boldsymbol{T}} \exp \left(s(t^{\prime}, a) / \tau\right)}\right],
\end{aligned}
\end{equation}
where $\boldsymbol{T}$ represents the set of text candidates. $f_{\boldsymbol{\theta}^{\boldsymbol{t}}}(\cdot)$, $f_{\boldsymbol{\theta}^{\boldsymbol{a}}}(\cdot)$ denote the text and audio encoders, respectively. This loss narrows the gap between semantically related text and audio in their representation space, which also helps the diffusion model to generate the joint distribution of text and audio. 


\begin{table*}
  \caption{Text-to-audio and audio-to-text retrieval performances in out-of-domain retrieval settings. “AudioCaps $\rightarrow$ Clotho” indicates that “AudioCaps” is the source domain (the training dataset) and “Clotho” is the target domain (the evaluation dataset).}
  \vspace*{-\baselineskip}
  \centering
  \begin{tabular}{c|cccccc}
    \toprule
    \multirow{3}{*}{Methods}
    & \multicolumn{6}{c}{AudioCaps $\rightarrow$ Clotho} \\
    \cline{2-7}
    & \multicolumn{3}{c}{Text-to-Audio} & \multicolumn{3}{c}{Audio-to-Text} \\
    \cline{2-7}
    & \textbf{R@1} & \textbf{R@5} & \textbf{R@10} & \textbf{R@1} & \textbf{R@5} & \textbf{R@10}\\
    \midrule
    ResNet38 (baseline) & 9.1$\pm$0.3 & 27.2$\pm$0.1 & 39.4$\pm$0.5 & 11.2$\pm$0.8 & 29.0$\pm$1.2 & 39.8$\pm$0.3\\
    CNN14 (baseline) & 8.3$\pm$0.2 & 25.9$\pm$0.4 & 37.5$\pm$0.7 & 9.2$\pm$0.4 & 26.2$\pm$0.3 & 38.1$\pm$0.4\\
    ResNet38 (Diff) & \textbf{13.1$\pm$0.3} & \textbf{31.4$\pm$0.5} & \textbf{43.3$\pm$0.4} & \textbf{15.4$\pm$0.2} & \textbf{33.9$\pm$0.4} & \textbf{44.3$\pm$0.1}\\
    CNN14 (Diff) & \textbf{12.2$\pm$0.2} & \textbf{30.4$\pm$0.4} & \textbf{41.9$\pm$0.1} & \textbf{14.3$\pm$0.5} & \textbf{30.5$\pm$0.3} & \textbf{42.8$\pm$0.7}\\   
  \bottomrule
\end{tabular}
\vspace*{-\baselineskip}
\end{table*}

\begin{table}\footnotesize
  \caption{Effect of the training strategy on the AudioCaps dataset.}
  \vspace*{-\baselineskip}
  \centering
  \begin{tabular}{c|ccc|ccc}
    \toprule
    \multirow{2}{*}{Method} & \multicolumn{3}{c|}{Text-to-Audio} & \multicolumn{3}{c}{Audio-to-Text} \\
    & \textbf{R@1} & \textbf{R@5} & \textbf{R@10} & \textbf{R@1} & \textbf{R@5}  & \textbf{R@10}\\
    \midrule      
    Gen & 26.2 & 62.8 & 76.7 & 31.9 & 66.7 & 78.6 \\   
    Dis & 33.9 & 69.7 & 82.6 & 39.4 & 72.0 & 83.9 \\
    Gen + Dis & \textbf{36.1} & \textbf{71.9} & \textbf{84.9} & \textbf{42.6} & \textbf{74.4} & \textbf{86.6} \\ 
  \bottomrule
\end{tabular}
\vspace*{-\baselineskip}
\end{table}

\begin{table}\footnotesize
  \caption{Effect of the diffusion steps on the AudioCaps dataset.}
  \vspace*{-\baselineskip}
  \centering
  \begin{tabular}{c|ccc|ccc}
    \toprule
    \multirow{2}{*}{Method} & \multicolumn{3}{c|}{Text-to-Audio} & \multicolumn{3}{c}{Audio-to-Text} \\
    & \textbf{R@1} & \textbf{R@5} & \textbf{R@10} & \textbf{R@1} & \textbf{R@5}  & \textbf{R@10}\\
    \midrule      
    10 & 35.9 & 71.6 & 84.6 & 42.2 & 74.1 & 86.2 \\   
    \textbf{50} & \textbf{36.1} & \textbf{71.9} & \textbf{84.9} & \textbf{42.6} & \textbf{74.4} & \textbf{86.6} \\ 
    100 & 36.0 & 71.8 & 84.4 & 42.3 & 74.3 & 86.1 \\
    200 & 35.9 & 71.2 & 84.1 & 42.2 & 73.9 & 86.1 \\
  \bottomrule
\end{tabular}
\vspace*{-\baselineskip}
\end{table}

\begin{table}\footnotesize
  \caption{Effect of the training batch size on the Clotho dataset.}
  \vspace*{-\baselineskip}
  \centering
  \begin{tabular}{c|ccc|ccc}
    \toprule
    \multirow{2}{*}{batch size} & \multicolumn{3}{c|}{Text-to-Audio} & \multicolumn{3}{c}{Audio-to-Text} \\
    & \textbf{R@1} & \textbf{R@5} & \textbf{R@10} & \textbf{R@1} & \textbf{R@5}  & \textbf{R@10}\\
    \midrule      
    16 & 16.2 & 37.7 & 51.4 & 18.7 & 40.0 & 52.2 \\   
    \textbf{24}  & \textbf{16.7} & \textbf{38.2} & \textbf{51.9} & \textbf{18.8} & \textbf{40.4} & \textbf{52.7}\\
    32 & 16.6 & 37.8 & 51.6 & 18.6 & 40.1 & 52.6 \\
  \bottomrule
\end{tabular}
\vspace*{-\baselineskip}
\end{table}

\section{Experiments}
\label{sec:exp}
\subsection{Datasets}
In our experiment, we leverage two publicly accessible datasets: AudioCaps \cite{kim2019audiocaps} and Clotho \cite{drossos2020clotho}. The AudioCaps dataset encompasses around 50K 10-second audio clips, which are sourced from AudioSet \cite{gemmeke2017audio}, the largest dataset for audio tagging. Within this collection, the training subset comprises 49,274 audio instances, with each having a unique human-derived caption. Both validation and test subsets include 494 and 957 audio instances respectively, and each instance is paired with five human-generated captions. Clotho v2, sourced from the Freesound platform \cite{font2013freesound}, features audio clips uniformly ranging from 15 to 30 seconds in length. Every instance in this dataset is coupled with five descriptive sentences. The distribution for training, validation, and testing subsets amounts to 3,839, 1,045, and 1,045 instances, respectively. 

\subsection{Training Details and Evaluation metrics}
In our work, we follow the same pipeline in \cite{mei2022metric} to train our network, where the batch size is set to 32 for the AudioCaps dataset and 24 for the Clotho dataset. The number of feature channels is 512 and the diffusion steps are set as $K$ = 50. We utilize BERT \cite{kenton2019bert} for text encoding, while employing ResNet38 and CNN14 from Pre-trained audio neural networks (PANNs) \cite{kong2020panns} for audio encoding. The training is divided into two phases. In the first phase, we train the feature encoder only using the contrastive loss. In the second phase, we freeze the feature encoder and only train the diffusion model, that is, we augment the cosine similarity with the similarity estimated by the diffusion model to get the final similarity. We repeat each experiment three times with different training seeds, presenting the mean and standard deviation of the metrics. We adopt the denoising diffusion implicit models (DDIM) \cite{song2020denoising} to estimate the distribution. During the inference phase, we consider both the distance between audio-text representations in their space and their joint probability. We evaluate using the Recall at rank k (R@k) metric, indicating the percentage of relevant targets in the top-k ranked results. The results of R@1, R@5, and R@10 are presented.

\subsection{Experimental Results}
In this part, we first compare the performance of our DiffATR with previous baselines in Table 1 and Table 2. We adopt either ResNet38 or CNN14 as the audio encoder on the AudioCaps and Clotho datasets. It is clear that our DiffATR brings significant boosts over different audio encoders for both datasets, which shows the effectiveness of our method. 

While existing ATR methods are typically trained and evaluated on the same dataset, we also explore out-of-domain ATR for deeper insights into the adaptability and generalization of our DiffATR framework. Here, we first train our model on one dataset (the “source”) and then measure its performance on another dataset (the “target”) that is unseen in the training. As shown in Table 3 and Table 4, we find that our DiffATR notably surpasses the corresponding discriminative approaches in out-of-domain ATR, which strongly demonstreates that our method is more generalizable and transferrable for the unseen data than conventional discriminative models.

\begin{figure}[t]
  \centering
  \includegraphics[width=1.0\linewidth]{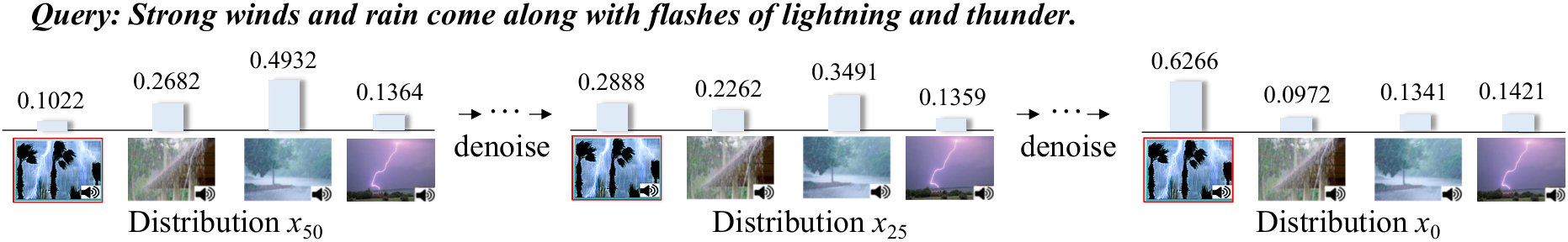}
  \caption{The visualization of the diffusion process of the probability distribution.}
  \vspace*{-\baselineskip}
\end{figure}

\subsection{Ablation Study}
In this part, we examine the impact of different training strategies, the number of diffusion steps, and the training batch size, utilizing ResNet38 as the audio encoder.

In Table 5, we compare different training strategies on the AudioCaps dataset. We observe that solely discriminative (Dis) training outperforms purely generative (Gen) training. The hybrid training method we present achieves the best results, confirming its capability to combine the strengths of both approaches. Table 6 studies the effect of diffusion step numbers. The results highlight that 50 steps is the sweet spot for the ATR task, surpassing the common range of 100-1000 steps used in audio generation tasks \cite{liu2023audioldm,yang2023diffsound}. We consider that this is due to the simpler probability distribution in retrieval compared to the generated audio signal distribution. Consequently, the ATR task demands fewer diffusion steps relative to generation-oriented tasks. Besides, we study the influence of the batch size on the Clotho dataset. As shown in Table 7, different training batch size settings would affect the results, but not too much, with the most favorable outcomes manifesting at a batch size of 24.

\subsection{Visualization}
In Figure 3, we provide the visualization of the diffusion process to better understand the diffusion process. The ground truth is marked in red blue, where we demonstrate the process from randomly initialized noise input ($x_{50}$) to the ultimate predicted distribution ($x_{0}$). The visualization shows that our approach incrementally reveals the relationship between text and audio.

\section{Conclusions}
\label{sec:conclusion}
In this paper, we present DiffATR, a generative diffusion-based framework for ATR. By explicitly modeling the joint probability distribution of text and audio, DiffATR optimizes the ATR network from both a generative and discriminative perspective to handle the intrinsic limitations of the current discriminative ATR paradigm. Experiments on the AudioCaps and Clotho datasets with superior performances, justify the effectiveness of our DiffATR. Remarkably, DiffATR also excels in out-of-domain retrieval scenarios without necessitating any modifications.

\bibliographystyle{IEEEtran}
\bibliography{mybib}

\end{document}